\documentclass[
superscriptaddress,
aip,
twocolumn,
showpacs,
showkeys,
reprint,
]{revtex4-1}

\usepackage{graphicx}
\usepackage{bm}

\begin{document}

\title{Radio-frequency dispersive detection of \\ donor atoms in a field-effect transistor}

\author{J. Verduijn$^*$}
\affiliation{Centre for Quantum Computation and Communication Technology, University of New South Wales, Sydney NSW 2052, Australia}

\author{M. Vinet}
\affiliation{CEA/LETI-MINATEC, CEA-Grenoble, 17 rue des martyrs, F-38054 Grenoble, France}

\author{S. Rogge}
\affiliation{Centre for Quantum Computation and Communication Technology, University of New South Wales, Sydney NSW 2052, Australia}

\date{\today}

\begin{abstract}
Radio-frequency dispersive read-out can provide a useful probe to nano-scale structures such as nano-wire devices, especially when the implementation of charge sensing is not straightforward. Here we demonstrate dispersive `gate-only' read-out of phosphor donors in a silicon nano-scale transistor. The technique enables access to states that are only tunnel-coupled to one contact, which is not easily achievable by other methods. This allows us to locate individual randomly placed donors in the device channel. Furthermore, the setup is naturally compatible with high bandwidth access to the probed donor states and may aid the implementation of a qubit based on coupled donors.
\end{abstract}



\maketitle

Radio-frequency techniques have proven to be a very powerful tool to read out semiconducting and superconducting mesoscopic devices. Conventionally it has been used to improve the detection sensitivity and speed up characterization of devices \cite{Schoelkopf:1998co} and thereby providing access to dynamics on much shorter time-scales compared to conventional direct-current techniques \cite{Ferguson:2006hg,Reilly:2007ig,Cassidy:2007ga}. More recently, however, it was realized that in some cases it can replace a charge detector by directly sensing the quantum capacitance of localized states \cite{Duty:2005iq,Petersson:wd}. In this way, read-out of a double quantum dot singlet-triplet qubit has recently been realized \cite{Petersson:2012cv}. Furthermore, radio-frequency multiplexing techniques can be utilized to selectively address multiple devices at the same time \cite{Stevenson:2002br}.
\par
Here, we study the utilization of a matching circuit attached to the gate of an etched nano-wire field-effect transistor. This allows us to map out donor states present below the device threshold, given that their charge transitions are responsive at the excitation frequency. The matching circuit consists of a resonant circuit which improves the power transfer between the device and the attached radio-frequency components. An advantage of this approach is that no optimization of the device lithography is needed, as it naturally accounts for the parasitic capacitance caused by, for example, the one-chip contact pads and bond wires \cite{Schoelkopf:1998co}. The setup and the device are very similar to the one used by Villes {\it et al.} \cite{Villis:2011bg}, who spectroscopically probed a dislocation in the gate contact of the device. We attach the matching circuit to the gate because localized states (e.g. donors) that are coupled to both leads as well as inter-channel transition are observable in this way \cite{Colless:2013fm,Frey:2012ce}. Furthermore, the present configuration is more easily scale-able for practical device applications; Since `gate-only' sensing is utilized there in no need for strongly tunnel coupled ohmic contacts \cite{Colless:2013fm}.
\par
%
\begin{figure}
\includegraphics[width=80mm]{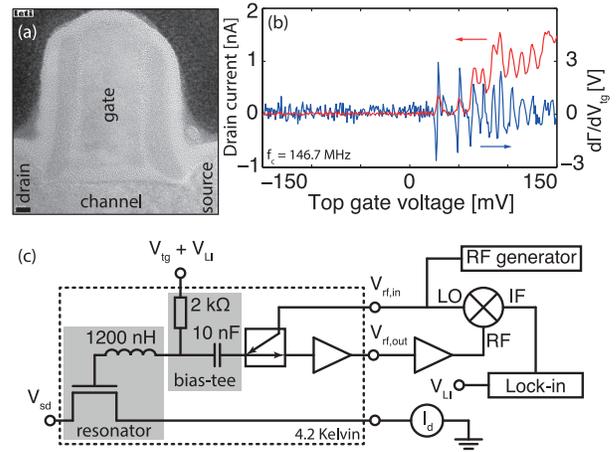}
\caption{\label{fig1} (a) A cross-sectional transmission electron microscope image of a device fabricated in the same batch as the devices measured in this work (scale bar on bottom right is 5 nm long). The device has been cut through over a line through the channel and gate between the source and drain contact (back gate not shown). (b) Measured drain current and dispersive signal versus gate voltage of device A (low doping) at a source-drain bias voltage of 100 $\mu$V. (c) Schematic diagram of the used homodyne detection setup.} 
\end{figure}
To demonstrate the ability to probe sub-threshold donor states, we measure devices with two different donor concentrations in the channel. The devices are field-effect transistors that consist of an nano-wire etched from a 20 nm thick silicon film on a silicon on insulator (SOI) wafer. The silicon film was first implanted with phosphorus dopants at a concentration of $10^{17}$ cm$^{-3}$ or $10^{18}$ cm$^{-3}$. A doped polycrystalline gate perpendicularly on top of the wire, isolated from the channel by a 5 nm thick layer of SiO$_2$, allows for gate control of the channel. The source and drain contact are formed in a self-aligned manner after the gate is deposited by implantation with arsenic donors at a concentration beyond the metal-insulator transition. Special care has been taken to make the junction with the channel as abrupt as possible thus minimizing diffusion of dopants from the contacts into the channel during later anneal steps. A substrate bias (back gate) can be used to set the electric field in the channel. The buried oxide between the channel and substrate is 145 nm thick. The electric field can tune single donor states and quantum dot-like states in our device \cite{Khalafalla:2007hj,Roche:2012dz,Verduijn:2013do}. For more details on the device structure and fabrication refer to Barral {\it et al.} \cite{Barral:2007co}.
\par
The data presented here has been collected using two different device, device A and B. Device A has channel dimensions of $50\times40\times20$ nm and has on average 2-6 randomly implanted donors in the channel ($10^{17}$ cm$^{-3}$) and device B is $140\times60\times12$ nm with on average 91-111 donors ($10^{18}$ cm$^{-3}$). As we will demonstrate below, the dispersive read-out will provide the same information as conventional current spectroscopy for device A, but for the much longer highly doped device B this is not the case. This device has many localized states that have only a small capacitive coupling to the source and drain contacts, but are strongly capacitively coupled to the gate. We will first discuss the data of device A with only a small number of donors in the channel and then move on to discuss device B and demonstrate a method to locate donor positions in the device channel using the dispersive read-out on this device. All the data presented has been taken with the devices submerged in liquid helium at 4.2 Kelvin.
\par
%
High frequency dispersive read-out is carried out by exciting the device through a custom-made bias tee with a low-power sinusoidal signal at a frequency, $f_c$, close to the resonance frequency of the matching circuit, with an amplitude of 35 $\mu$V (-76 dBm) \footnote{The amplitude of the signal impinging on the gate is enhance by the Q-factor, i.e. $V_{g,rf}=QV_{ex}$. Therefore we estimate the radio-frequency excitation amplitude is $\sim1$ mV when we tune the frequency exactly to the resonance frequency of the matching circuit.}. The matching circuit consists of an $L=1200$ nH chip inductor \footnote{Coilcraft 1206CS-122X$\_$LB SMD chip inductor} which provides cancelation of the device reactance at a frequency $f_0=1/2\pi\sqrt{LC}$ \footnote{The reactance of the device and carrier is mostly capacitive at the carrier frequency of 147 MHz}. The measured effective total capacitance, $C$, is about 0.90 pF, resulting in a resonance frequency of about 147 MHz. Since the dispersive read-out is loosely speaking only sensitive to processes that can respond to the radio-frequency signal at frequencies that are similar or larger than $f_c$, we choose to use a relatively large inductance to obtain a lower resonance frequency. We tested several device with different bond pad geometries as well as bond wire length on the same custom-made chip carrier, but always found approximately the same capacitance. This indicates that the contribution of the parasitic capacitance of the chip carrier dominates. The loaded quality factor of the matching circuit is measured to be $\sim31$ at 4 Kelvin \footnote{The Q-factor is mainly limited by external losses due to the 50 $\Omega$ load of the amplifier. We estimate the internal Q-factor to be about 47, based on a Spice simulation with the actual board and device parameters and a realistic model for the chip inductor.}. 
\par
\begin{figure}
\includegraphics[width=80mm]{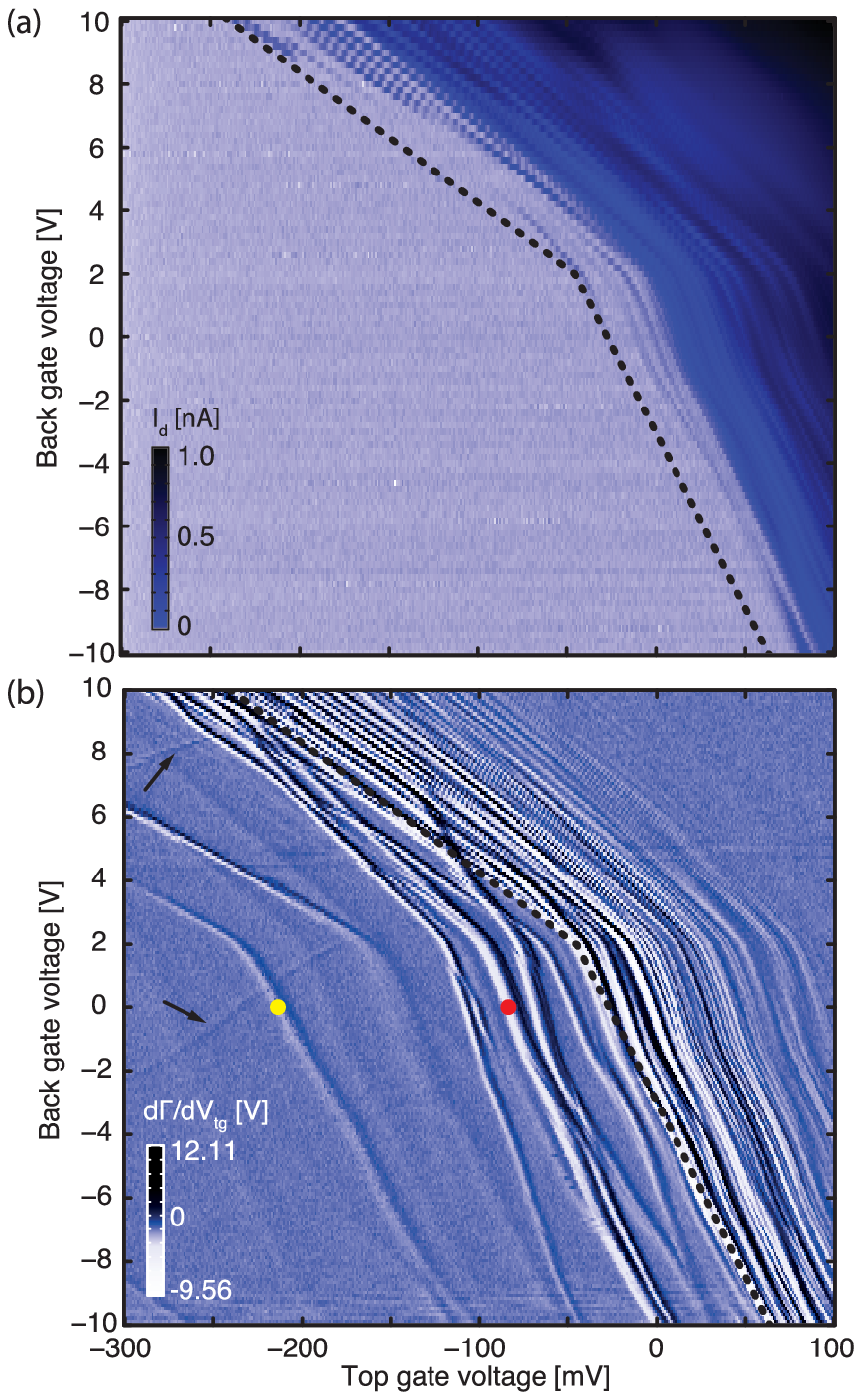}
\caption{\label{fig2} (a) The measured drain current as a function of top gate voltage and back gate voltage at 100 $\mu$V source bias voltage of device B. (b) The dispersive signal of the same device from the lock-in amplifier at an excitation frequency of 147.7 MHz. The dot (yellow and red) correspond to the same voltages as the dots in figure \ref{fig3}. The black arrows indicate diagonal resonances not tunnel coupled to the source and drain contacts. A black dashed line is drawn in (a) and (b) as guide to the eye to compare the device threshold in the different plots.} 
\end{figure}
The back-reflected signal is separated from the incoming signal \footnote{We use a Mini-Circuits ZEDC-15-2B directional coupler} and amplified by about 19 dB with a low-noise amplifier \footnote{Mini-Circuits ZX60-33LN-S+ from which we removed the supply voltage protection.} close to the device, see figure \ref{fig1}. The output signal of the cold amplifier is further amplified at room temperature \footnote{Two cascaded Mini-Circuits ZFL-1000LN+} and then down-converted using a mixer \footnote{Mini-Circuits ZEM-2B} in a homodyne detection scheme. A lock-in amplifier is used to improve the signal-to-noise by modulating the dc input of the bias tee (the gate) at a frequency of 269 Hz with an amplitude of 0.7 mV. The down-converted intermediate frequency (IF) signal is fed into the lock-in amplifier which is read out by a data acquisition system. The output signal of the lock-in is the (uncalibrated) derivative of the reflection coefficient with respect to the gate voltage, $\text{d}\Gamma/\text{d}V_{tg}$. Here, $\Gamma$ is the reflection coefficient given by $\Gamma=(Z-Z_0)/(Z+Z_0)$, with $Z_0=50$ $\Omega$ the impedance of the radio-frequency circuit and $Z$ the complex impedance of matching circuit. This setup provides a very sensitive probe to changes in the effective impedance of the gate of our device. Note that our scheme differs from the commonly used read-out of radio-frequency quantum point contacts and single electron transistors. In contrast to these device, where the response is mostly due to a change in device resistance \cite{Roschier:2004hf}, our gate has a practically infinite resistance.
\par
To establish the correspondence between the drain current signal and the dispersive signal from the device, $\text{d}\Gamma/\text{d}V_{tg}$, we plot both signals measured using device A in figure \ref{fig1} (b). Since we effectively measure the derivative of the dispersive signal with respect to the gate voltage, and therefore, a single Coulomb blockade peak shows up as an anti-resonance. We find that the dispersive signal is strongest for the first few Coulomb blockade resonances. 
The derivative signal with respect to the gate voltage the contrast in the signal reduces because the Coulomb peaks vanish at higher gate voltage. For this short-channel device with only very few donors present, we do not observe any resonances besides the Coulomb blockade resonances also visible in the current signal. This situation is very different for device B which has a longer channel with many more donors in it.
\par
To map out the device channel we sweep the top gate voltage as, $V_{tg}$, well as the back gate voltage, $V_{bg}$, of device B and record the drain current and dispersive signal, see figure \ref{fig2} (a) and (b). In contrast to device A, the drain current of this device shows Coulomb blockade resonances beyond a line that corresponds to condition where interface-well states are induced at the top gate side of the channel for back gate voltages $\gtrsim2$ V and at the back gate side for $\lesssim2$ V \cite{Roche:2012dz}. The slope $\beta=\text{d}V_{tg}/\text{d}V_{bg}$ of a resonance quantifies the relative capacitive coupling to the top gate and back gate contacts. Because the capacitive coupling to the back gate is stronger (larger $\beta$) when these states reside at the back side of the channel as compared to states near the top gate (smaller $\beta$), a distinctive kink in the low temperature threshold voltage develops near $\sim2$ Volt on the back gate \cite{Verduijn:2013do}. Since the channel of this device is too long to support direct resonant tunneling through sub-threshold donor states, the current is blocked at smaller gate voltages. The dispersive signal, however, shows a very rich structure of resonances. We have performed the same experiment for the device A and found no additional resonances. Therefore, we attribute these resonances to the presence of the phosphorus donors in the channel. Surprisingly, we observe resonances at top gate voltages as low as -300 mV even though one may expect that the donors in the channel are all ionized as they are located far above the chemical potential of the source and drain contacts. These resonances are due to donors that are located in the channel close to the contacts and therefore only weakly coupled to the gate. This is also apparent from the fact that the resonances at lower gate voltages are generally much wider than the ones present above the threshold line.
\par
The value of $\beta$ for the resonances in figure \ref{fig2} (b) provide information about the location and nature of the states. For example, when the slope $\beta$ is shallower (larger) than the slope of the interface-well states around and above threshold we conclude that the state must be present in the bulk of the channel. Note that when $\beta<0$ (anti-diagonal lines), the charge transfer takes place to the source and drain contacts, which are kept at a constant voltage. But there are also several resonances visible for which $\beta > 0$ (diagonal lines), for example the ones denoted by the black arrow in figure \ref{fig2} (b). A resonance with a positive slope can be explained as a transition between two states both localized in the channel and capacitively coupled to the gates similar to inter-dot transitions as observed in double quantum dots \cite{VanderWiel:2002tz,Petersson:wd,Basset:2013fn}, but a more likely explanation in this case are localized states in the gate. This is because for transitions between localized state inside the channel, the lines are expected to terminate (or at least change slope) when the overall charge state of the system changes. Since we do not see the diagonal lines terminate, even when $V_g$ is as low as -300 mV, we believe that disordered metallic islands in the polycrystalline silicon gate are the cause. In the following, we focus on the dispersive signal from device B as a function of source-drain bias voltage.
\par
\begin{figure}
\includegraphics[width=80mm]{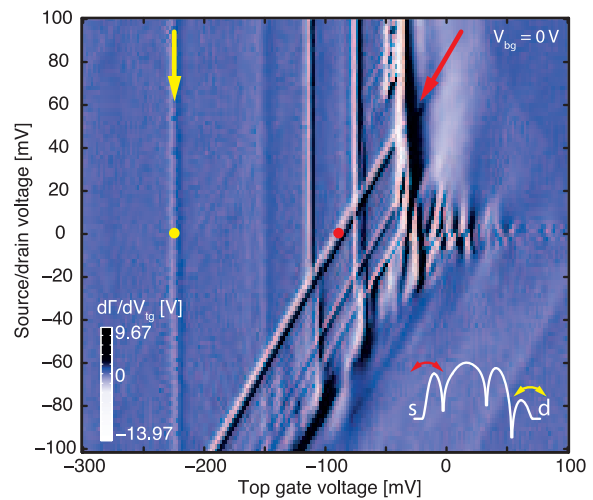}
\caption{\label{fig3} (a) Measured dispersive signal at under the same conditions as figure \ref{fig2} at zero back gate voltage (on device B). The processes corresponding to the resonances denoted by the yellow and red arrows are schematically depicted in the lower right corner.} 
\end{figure}
To illustrate the advantages of dispersively probing localized states using a gate, we plot the dispersive signal as a function of source-drain voltage and gate voltage in figure \ref{fig3}. We observe resonances associated with localized states on the source side as well as on the drain side simultaneously. A vertical line, as for example the one denoted by the yellow arrow in figure \ref{fig3}, means a resonance associated with charge transfer to the grounded drain contact whereas a diagonal line (red arrow) is a state on the source side of the channel. This information is essentially the relative capacitive coupling to the source in drain contacts, which depends on the geometric location of the donor inside the channel. By combining this knowledge with the capacitive coupling to the top gate and back gate as can be obtained from figure \ref{fig2} (b), we can determine an approximate location for each donor that is observed in the dispersive signal.
\par
In devices in which a charge sensor has proven to be difficult to implement dispersive read-out of the device can often fill in this gap. Such devices include etched nano-wires devices such as studied here as well as devices made of bottom-up grown nano-wire and carbon nano-tubes. The etched nano-wire devices studied here has a geometry that is especially suitable for this `gate-only' read-out technique, since the gate is tightly wrapped around the wire. This makes the capacitive coupling of the gate to the localized states large compared to the other contacts. In devices similar to the ones in this work, we have observed ratio of the gate capacitance over the total capacitance up to $\sim0.9$. In a highly optimized structure with a superconducting strip-line resonator with a high loaded quality factor dispersive read-out of a double quantum dot has recently even proven to be superior to charge sensing using a quantum point contact \cite{Basset:2013fn}. However, here we focus on a setup using a chip inductor with an inherently low quality factor. This has as an advantage that the detection bandwidth can be much larger. Provided that the signal to noise ratio is sufficient, this can make fast read-out of quantum states possible. In the current setup the parasitic capacitance of the used circuit board is dominating, therefore, improvements can be made by re-designing the circuit board. Other future work to improve the signal to noise ratio may include the implementation of amplitude modulation techniques which allow the use of radio-frequency amplifiers and detectors \cite{Reilly:2007ig}.
\par
In conclusion, we have demonstrated `gate-only' read-out of donor states in a silicon nano-wire field effect transistor. This enables the detection of localized states that are only tunnel coupled to either the source or the drain contact as well as the identification of metallic localized states in the gate. By looking at the relative capacitive coupling to the contacts we can determine a relative position of individual donors in channel. The results presented here may also be useful to read-out inter-donor charge transitions which can preserve spin coherence and thereby be used for spin-based quantum device applications.
\begin{acknowledgements}
The devices have been designed and fabricated by the AFSiD Project partners, see http://www.afsid.eu. The authors thank Junxi Duan and Pablo Asshoff for their assistance in conducting the experiment and Joost van der Heijden for his help with the design of the used printed circuit board chip carrier. J.V. thanks Tim Duty for the motivating discussions about the radio-frequency technique. This work was supported by the EC FP7 FET-proactive NanoICT projects AFSiD (214989) and TOLOP (318397) and by the Australian Research Council (ARC) Centre of Excellence for Quantum Computation and Communication Technology (CE110001027). S.R. acknowledges an ARC Future Fellowship (FT100100589).
\end{acknowledgements}
$^*$Corresponding author: a.verduijn@unsw.edu.au

\end{document}